\documentclass{phb-proc4-auth}

\usepackage{graphicx}
\usepackage{amssymb}

\newlength{\piclen}
\begin{document}
\begin{frontmatter}


\journal{SCES '04}


\title{Two Aspects of the Mott-Hubbard Transition in Cr-doped V$_2$O$_3$}

%
%
%
%
%
%

\author[MPI]{K. Held}
\author[MI]{J. W. Allen}
\author[E]{V. I. Anisimov}
\author[A]{V. Eyert}\linebreak
\author[A]{G. Keller}
\author[K]{H.-D. Kim}
\author[MI]{S.-K. Mo}
\author[A]{D. Vollhardt}

%
 
\address[MPI]{Max-Planck-Institut f\"ur
Festk\"orperforschung, 70569 Stuttgart, Germany}
\address[MI]{Randall Laboratory of Physics, University of
Michigan, Ann Arbor, MI 48109, USA}
\address[E]{Institute of Metal Physics, Ekaterinburg GSP-170, Russia}
\address[A]{Institut f\"ur Physik, Universit\"at Augsburg,
         86135 Augsburg, Germany}
\address[K]{Pohang Accelerator Laboratory, Pohang 790-784, Korea}
%
%
%
%


%
%
%
%


\begin{abstract}
The combination of bandstructure theory in the local density approximation with dynamical mean field theory was
recently successfully  applied to V$_2$O$_3$ -- a material which undergoes the famous Mott-Hubbard metal-insulator transition upon Cr doping. The aim of this short paper is to emphasize two aspects of our recent results: (i) the filling of the Mott-Hubbard gap with increasing temperature, and (ii) the peculiarities of the Mott-Hubbard transition in this system which is not characterized by a divergence of the effective mass for the  $a_{1g}$-orbital.
\end{abstract}

%
%



\end{frontmatter}

%
%
%
%
%

{\it Introduction --}
A paradigm of electronic correlations is the Mott-Hubbard metal-insulator transition,
triggered by an increasing ratio of Coulomb interaction over bandwidth ($U/W$). 
Experimentally,
it is realized  in paramagnetic V$_2$O$_3$ upon Cr-doping or decreasing 
pressure.  Theoretically, 
the combination of the local density approximation and dynamical mean field theory
(LDA+DMFT\cite{LDADMFT}) was successfully employed to this system recently \cite{V2O3theory},
stimulating new experiments \cite{mo}. Here we restrict ourselves to two fascinating  
aspects of the transition: (i) the filling of the Mott-Hubbard gap with increasing temperature \cite{mo}
and (ii)  the peculiarities of the Mott-Hubbard transition
due to the inequivalence of the $a_{1g}$- and $e_g$-orbitals \cite{V2O3theory}.

\begin{figure}[b]
\vspace{-.4cm}

\hspace{-.3cm}\includegraphics[clip=true,width=1.07\piclen]{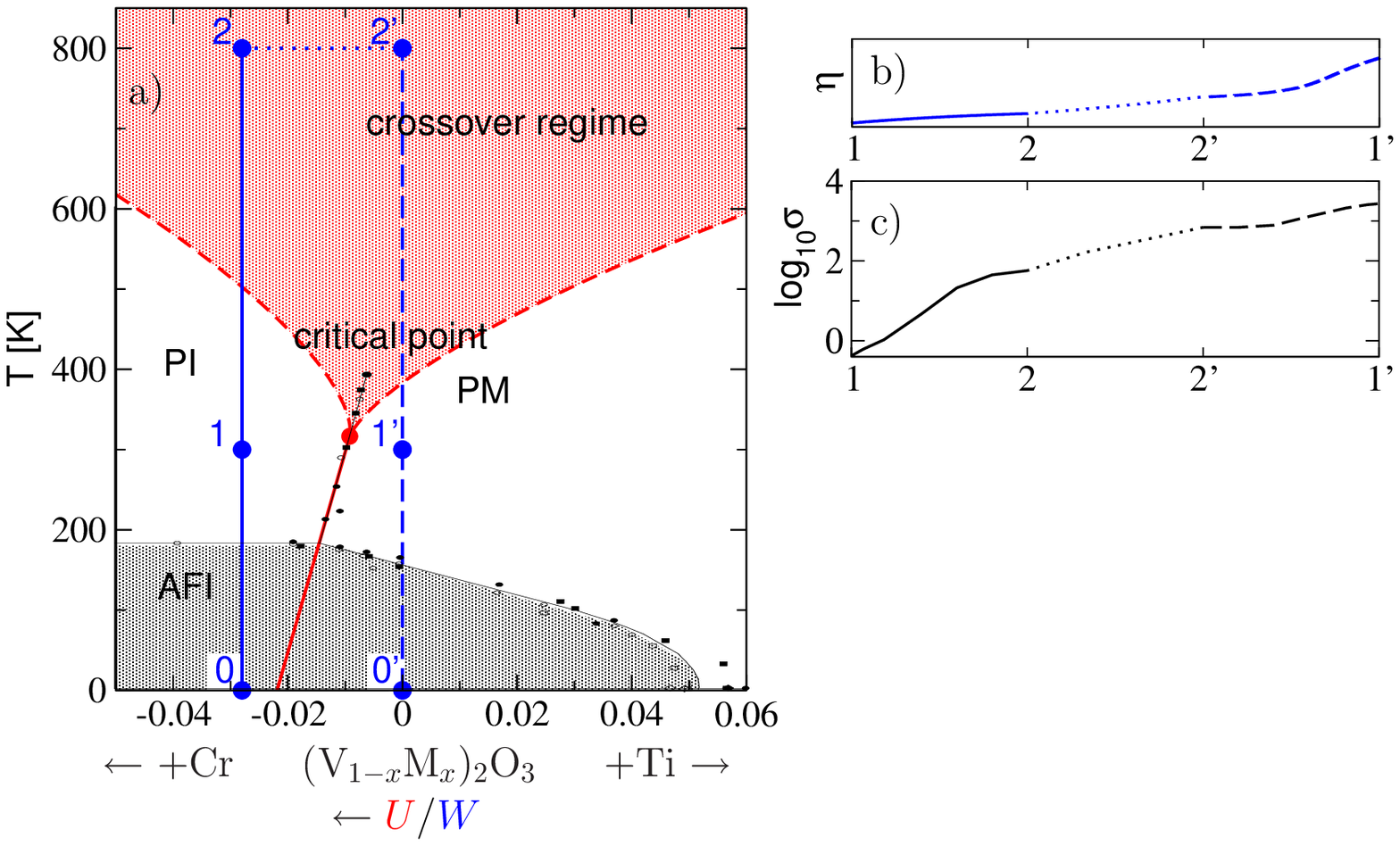}\\
\vspace{-1.1cm}

\caption{a) Experimental (black/white) and  DMFT (colors) phase diagram 
 of ${\rm V_2O_3}$;
b) order parameter of the DMFT Landau theory and 
c) logarithm of the experimental conductivity (in units of
$[\Omega {\rm cm}]^{-1}$) along the paths of a)
[reproduced from \cite{mo} with c) read off
from \cite{exprho}].}
\label{figPD}
\end{figure}

{\it (i) Filling of the Mott-Hubbard gap with increasing temperature --}
A feature genuine to a Mott-Hubbard insulator is that the zero temperature ($T$) Mott-Hubbard gap
is filled with spectral weight when  $T$ is increased.
In contrast, for  a semiconducting or band insulator
a gap stays a gap with increasing $T$;
only some electrons are statistically transferred across the gap because of
finite $T$. How can we understand this non-intuitive feature of the Mott-Hubbard insulator?
 Let us follow the two
 paths  for  metallic V$_2$O$_3$ and insulating (V$_{0.972}$Cr$_{0.028}$)$_2$O$_3$
in the phase diagram Fig.\ \ref{figPD}a.
If we consider a paramagnet instead of the antiferromagnetic insulator  (AFI)
at zero temperature, the paramagnetic metal (PM) 
is a Fermi liquid with a central quasiparticle resonance
(point 0') and the insulator (PI) is gapped (point 0).
This picture changes only slightly if we increase $T$ to
points 1' and 1. 
 When T is however further increased (to point 2') one enters
the crossover regime, the quasiparticle peak fades away, and
the lifetime becomes too short to speak of a quasiparticle
peak anymore.
Since the insulator and the metal are not distinct
phases in the crossover regime,
we see the Mott-Hubbard gap being filled with incoherent  (short lifetime)
spectral weight as one goes from point 1 to 2. This weight may be interpreted as a smeared out
quasiparticle peak.

\begin{figure}[tb]

\begin{center}
\vspace{-.3cm}

\includegraphics[clip=true,width=0.75\piclen]{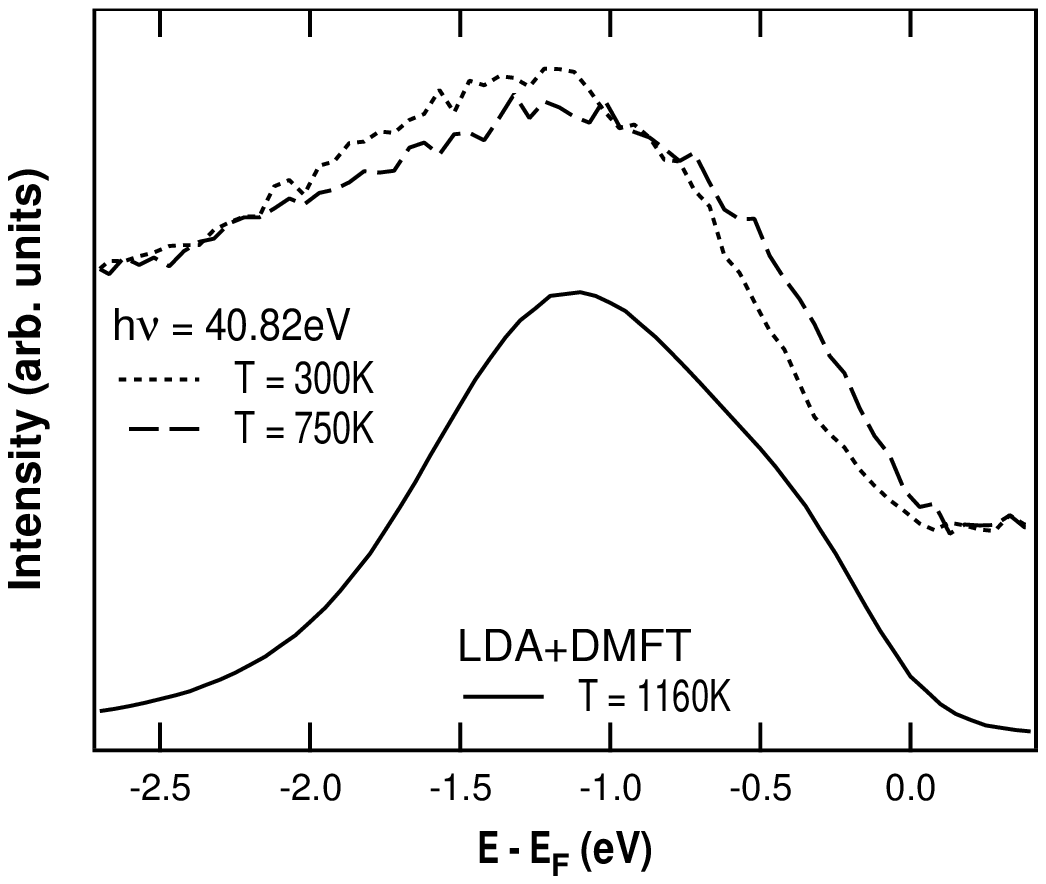}
\end{center}
\vspace{-.1cm}

\caption{Photoemission spectra showing the transfer of
spectral weight into the  Mott-Hubbard gap with
increasing $T$  and  LDA+DMFT prediction  \cite{V2O3theory} [reproduced from \cite{mo}].}
\label{figPES}
\end{figure}

In Fig.\ \ref{figPD}b, the calculated \cite{mo}
 order parameter $\eta$  of the
DMFT Landau theory \cite{DMFTLandau}
is plotted. Since $\eta$
corresponds to the spectral weight at the Fermi energy,
it shows in detail
the reduction of the metallic quasiparticle peak and the filling of the Mott-Hubbard gap along
the lines from point 1' to 2' and  from 1 to 2, respectively.
As was pointed out in \cite{mo}, this behavior is indeed seen in the
experimental  conductivity (Fig.\ \ref{figPD}c). Mo {\em et al.} \cite{mo}
particularly observed for the first time the filling of the 
Mott-Hubbard gap with increasing $T$ by photoemission spectroscopy, see Fig.\ \ref{figPES},
ascertaining the DMFT prediction.

\begin{figure}[tb]
\includegraphics[clip=true,width=0.99\piclen]{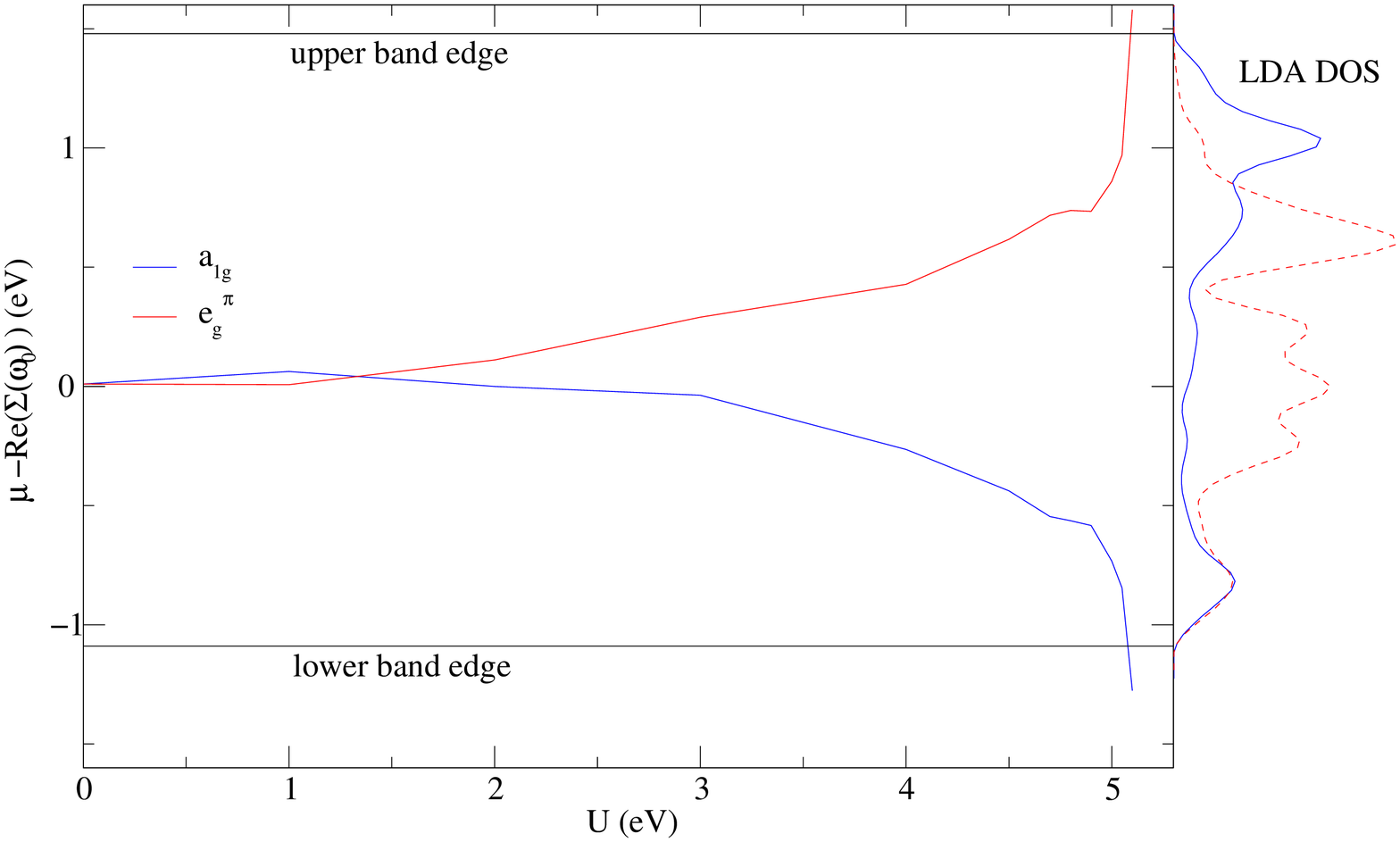}
\caption{The LDA+DMFT effective chemical potential $\mu -\Re\Sigma_m (\omega_0)$,
for the $a_{1g}$- and $e_g^\pi$-orbitals (taken 
at the lowest Matsubara frequency $\omega_0$)
moves out of the band edges of the LDA DOS  at the critical $U$ for the Mott-Hubbard 
transition [reproduced from \cite{V2O3theory}].}
\label{figmueff}
\end{figure}

{\it (ii) Peculiarities of the Mott-Hubbard transition --}
Let us  now turn to a different aspect of the Mott-Hubbard transition which stems from
the orbital degrees of freedom. 
Within the Fermi-liquid metallic phase of V$_2$O$_3$, the 
 Green function is given by
\begin{eqnarray}
G_m(\omega)&\!=\!&\int {\rm d}\epsilon\;
\frac{Z N_{m}^{0}(\epsilon )}{\omega+Z(\mu -\Re\Sigma_m (0)- \epsilon)}
\label{FL}
\end{eqnarray}
at low frequencies $\omega$. Here, $N_{m}^{0}(\epsilon)$ is the non-interacting (LDA) density of states (DOS)
for $m=e_g^\pi$ or $m=a_{1g}$, $\Sigma_m$
denotes the self energy and
 $Z_m=(1\!-\!\partial \Re\Sigma_m(\omega)/{\partial \omega}|_{\omega=0})^{-1}$
the quasiparticle weight.
In the one-band Hubbard model, the Mott-Hubbard transition
is characterized by  $Z\rightarrow 0$ at the critical value 
of $U$, or equivalently a divergence of the effective mass.
The width of the quasiparticle peak goes to zero.
Keller {\em et al.} \cite{V2O3theory} showed  in their LDA+DMFT calculation
 that the Mott-Hubbard transition
in  V$_2$O$_3$ is actually different. While   $Z_{e_g^\pi}\rightarrow 0$ for the
two  $e_g^\pi$-bands, $Z_{a_{1g}}\!\not\rightarrow\! 0$  for the  $a_{1g}$-band at the Mott-Hubbard transition.
How can the  $a_{1g}$-band then become insulating?
Looking at Eq.\ (\ref{FL}), we see that there is another
possibility consistent with the Fermi liquid metallic phase. 
The height of the quasiparticle peak is given by
$N_m(\mu -\Re\Sigma_m (0))$. 
As is shown in Fig.\  \ref{figmueff}, 
 $\mu -\Re\Sigma_m (0)$ moves  out of the non-interacting DOS at the Mott-Hubbard transition. Hence,
the height of the quasiparticle peak goes to zero.
For  the $a_{1g}$  quasiparticle peak, the height goes to zero at fixed  
width (which is already strongly reduced
compared to $U=0$).
For the  $e_{g}^\pi$-orbitals,  the transition is characterized by a combined shrinking of width and height.

%
%
%
%

We acknowledge support of the DFG through the Emmy Noether program
and SFB  484,
U.S. NSF at the University of
Michigan (UM) (Grant No.~DMR-03-02825), and KOSEF through eSSC
at POSTECH.

%
%
%
%



\begin{thebibliography}{00}


\bibitem{LDADMFT} V.\ I.\ Anisimov {\em et al.}, 
A.\ I.\ Poteryaev, M.\ A.\  Korotin, A.\ O.\ Anokhin, and G.\ Kotliar, 
J.\ Phys.\ Cond.\ Matter {\bf 9}, 7359 (1997);
A.\ I.\ Lichtenstein and M.\ I.\ Katsnelson, Phys.\ Rev.\ B {\bf 57},
6884 (1998); for a review see  K.\ Held {\em et al.},
Psi-k Newsletter \#56, 65 (2003) 
\verb![!http://psi-k.dl.ac.uk/newsletters/News\_56/Highlight\_56.pdf\verb!]!.


\bibitem{V2O3theory} K.\ Held {\em et al.}, 
Phys.\ Rev.\ Lett.\ {\bf 86}, 5345 (2001);
 G.\ Keller {\em et al.}, 
cond-mat/0402133.

\bibitem{mo} S.-K.\ Mo {\em et al.},
Phys.\ Rev.\ Lett.\ {\bf 90}, 186403 (2003); Phys.\ Rev.\ Lett.\ in print [cond-mat/0403094].

\bibitem{exprho} H. Kuwamoto, J. M. Honig, and J. Appel, Phys. Rev. B {\bf 22}, 2626 (1980).

\bibitem{DMFTLandau} G.\ Kotliar, E.\ Lange, and M.\ Rozenberg, Phys.\ Rev.\ Lett.\
{\bf 84}, 5180 (2000).



\end{thebibliography}
\end{document}